\documentclass[referee]{raa}  
\usepackage{placeins}
\usepackage{graphicx,times}
\usepackage{natbib}
\usepackage{amssymb,amsmath}
\bibpunct{(}{)}{;}{a}{}{,}

\usepackage{enumitem}
\usepackage{float}
\usepackage{subcaption}
\graphicspath{{images/}}

\usepackage[hidelinks,pagebackref=true]{hyperref} 

\begin{document}

\title{The Quenching Mechanisms of Field Dwarf Galaxies}

\volnopage{Vol.0 (20xx) No.0, 000--000} 
\setcounter{page}{1}                    

   \author{Zuoyu Wu \inst{1,2}
    \and  Yougang Wang\inst{1,2,3}
    \and Shihong Liao \inst{4}  
          }

   \institute{
   State Key Laboratory of Radio Astronomy and Technology, National Astronomical Observatories, CAS, A20 Datun Road, Chaoyang District, Beijing, 100101, P. R. China; {\it wangyg@bao.ac.cn}\\
        \and
    School of Astronomy and Space Science, University of Chinese Academy of Sciences, Beijing 100049, P. R. China\\
        \and
    Key Laboratory of Cosmology and Astrophysics (Liaoning) \& College of Sciences, Northeastern University, Shenyang 110819, China\\
        \and
    Key Laboratory for Computational Astrophysics, National Astronomical Observatories, CAS, A20 Datun Road, Chaoyang District, Beijing, 100101, P. R. China\\}

\abstract{ Isolated dwarf galaxies are intrinsically faint and difficult to detect. The limited sample size makes it challenging to observationally constrain the physical mechanisms that quench their star formation. To disentangle the quenching mechanisms of isolated dwarfs, we identify a non-negligible population of such galaxies in the TNG50 simulation. In addition to the previously discovered ``backsplash" galaxies that were quenched by environmental effects when they were once satellites in more massive halos, we find another primary quenching channel in a population of galaxies whose star formation is suppressed by excessively strong gas outflows that prevent the gas from cooling and collapsing to form stars. We further demonstrate that these outflows are highly likely driven by stellar feedback and predominantly occur in high-gas-fraction dwarfs, which within our studied stellar mass range ($10^7$--$10^{9.5},M_\odot$) are always located toward the low-mass end.
}

    \authorrunning{Wu \& Wang}
    \titlerunning{Quenching of Field Dwarf Galaxies}
    \maketitle

\section{Introduction}

Dwarf galaxies (with stellar masses $M_\star \lesssim 10^{9.5}\,M_\odot$) are the most numerous population of galaxies in the Universe 
\citep[e.g.,][]{2012MNRAS.421.1007K, Moustakas_2013, 2017MNRAS.470..283W, 2025ApJ...994..231K} and provide unique laboratories for studying the regulation of star formation 
in the low-mass regime. 
Their shallow gravitational potentials make them particularly susceptible to internal and external processes that can 
readily alter their gas content and star formation efficiency 
\citep[e.g.,][]{10.1093/mnras/stw2982, Zheng2025}. 
Because these systems are typically too small to host active galactic nuclei (AGN), 
their evolution is governed primarily by stellar feedback, gas accretion, and environmental interactions, 
making them ideal testbeds for isolating the effects of stellar feedback on galaxy evolution. 
However, their intrinsically low luminosities and diffuse gas distributions make them observationally challenging \citep{Bidaran_2025}, 
especially for systems located in low-density environments, where environmental signatures are weak and direct 
measurements of gas dynamics are rare \citep{Weisz_2014, Gallart_2015}.

In the conventional picture of galaxy evolution, dwarf galaxies quench their star formation primarily through 
environmental mechanisms such as ram-pressure stripping, tidal interactions, or strangulation 
\citep{1972ApJ...176....1G, 10.1046/j.1365-8711.1999.02715.x, 10.1111/j.1365-2966.2012.21188.x}. 
These effects are strongest for satellite dwarfs orbiting within the virial radius of more massive hosts, 
where the surrounding hot gas can efficiently remove the cold interstellar medium (ISM). 
Such mechanisms naturally explain the predominance of quenched dwarf spheroidals around the Milky Way and Andromeda 
\citep[e.g.,][]{2018MNRAS.478..548S, 10.1093/mnras/stad2576}. 
There is an abundance of field dwarf galaxies that reside in isolation from massive hosts, and they are predominantly gas-rich and actively star-forming~\citep{Geha_2012}. Consequently, quenched systems in the field appear to be rare, although a small but non-negligible population has been reported in both observations and simulations.
These so-called isolated quenched dwarfs challenge the canonical, environment-centric view of quenching, 
suggesting that star formation can cease even in the absence of strong external forces.

Several physical mechanisms have been proposed to explain this puzzling population. 
One possibility is the backsplash scenario: galaxies that were formerly satellites of more massive hosts but are now located outside the hosts' virial radii \citep[e.g.,][]{2005MNRAS.356.1327G, 2011MNRAS.412..529K}. Another explanation is cosmic web stripping, whereby dwarf galaxies moving rapidly through 
filamentary or sheet-like structures of the large-scale cosmic web experience ram pressure from the diffuse intergalactic medium, 
leading to the removal or heating of their gas reservoirs \citep[e.g.,][]{2013ApJ...763L..41B}. These two mechanisms were recently discussed in detail in  \cite{benavides2025environmental}.
A third complementary scenario invokes \textit{internal quenching}, 
in which sustained stellar feedback drives powerful outflows that gradually deplete the cold gas or prevent its re-accretion. These processes may act simultaneously or sequentially, and their relative importance likely depends on both 
stellar mass and environment.

While the majority of quenched dwarfs are gas-poor systems, recent simulation results indicate that a subset of isolated quenched dwarfs retain substantial amounts of gas despite their lack of ongoing star formation \citep{rey2020edge}. The existence of such gas-rich but quenched systems raises an important question: 
how can galaxies with large gas reservoirs fail to form stars? Possible explanations include inefficient cooling \citep{10.1093/mnras/256.1.43P}, highly disturbed gas kinematics \citep{tamburro2009driving, stilp2013global}, or feedback-driven outflows 
that maintain the gas in a warm or diffuse state \citep{hopkins2014galaxies}. 
Characterizing the gas distribution and outflow properties of these systems is therefore crucial to understanding 
the final stages of quenching in low-mass galaxies.

In this work, we investigate the gas content and outflow properties of isolated quenched dwarf galaxies in the 
TNG50 cosmological hydrodynamical simulation \citep{2019MNRAS.490.3234N, 10.1093/mnras/stz2338}. 
We focus specifically on those dwarfs that retain a significant cold or total gas mass at $z=0$, 
and compare them with gas-poor quenched counterparts to explore how gas kinematics and outflow efficiency differ 
between the two populations. Our goal is to reveal whether high gas fractions in quenched dwarfs correspond to ongoing feedback-driven outflows, 
and how these processes contribute to the cessation of star formation. 
This study thus provides new insight into how feedback and environmental mechanisms jointly shape the late-time 
evolution of dwarf galaxies in isolation. The paper is organised as follows: Section 2 describes the simulation and sample selection, Section 3 presents the main results, and Section 4 discusses the physical interpretation and implications. 
\section{Data and Methods }
\subsection{IllustrisTNG Cosmological Simulations }
%

    We conduct analysis using the IllustrisTNG simulation \citep{2018MNRAS.480.5113M, 2018MNRAS.477.1206N, 2018MNRAS.475..624N, 2018MNRAS.475..648P, 2018MNRAS.475..676S}. The IllustrisTNG project is a series of cutting-edge, large-scale cosmological magneto-hydrodynamical simulations designed to study galaxy formation. Its goal is to uncover the physical mechanisms that shape how galaxies form and evolve over cosmic time, ultimately explaining the structures we observe in the universe today. The simulations are performed using the moving mesh AREPO code \citep{2010MNRAS.401..791S}. The adopted galaxy formation subgrid model is described in detail in \citet{Weinberger2017} and \citet{Pillepich2018}.
    The project encompasses three main simulation runs—TNG50, TNG100, and TNG300—which cover different cosmic volumes and resolutions. These simulations adopt a flat $\Lambda$CDM cosmology with parameters based on the \cite{2016A&A...594A..13P}: $h = 0.6774$, $\Omega_{\mathrm{m}} = 0.3089$, $\Omega_{\Lambda} = 0.6911$, $\Omega_{\mathrm{b}} = 0.0486$, $n_{\mathrm{s}} = 0.9677$, and $\sigma_8 = 0.8159$.
Among the three simulation volumes, TNG50-1 \citep{2019MNRAS.490.3234N, 10.1093/mnras/stz2338}—spanning $35\, \mathrm{Mpc}\, h^{-1}$ per side and including $2160^3$ dark matter particles and an equal number of initial gas cells—is particularly well-suited for studying dwarf galaxies in the present-day (z = 0) universe. This is due to its high resolution, with a baryonic mass resolution of approximately $m_{\mathrm{baryon}} \sim8.5 \times 10^4\, M_\odot$ and a dark matter particle mass of $m_{\mathrm{DM}} \sim 4.5 \times 10^5\, M_\odot$. Such a mass resolution is sufficient to reliably resolve galaxies containing about 100 stellar particles. The $z=0$ Plummer equivalent gravitational softening length of dark matter and star particles is 288 pc, and the minimum of the adaptive gas gravitational softenings is 72 pc. Halos and subhalos (or galaxies) in the TNG simulations are identified using the friends-of-friends \citep[FoF,][]{1985ApJ...292..371D} and SUBFIND algorithms \citep{2001MNRAS.328..726S, 2009MNRAS.399..497D}.
\begin{figure}[htbp]
    \centering
    \includegraphics[width=1.0\linewidth]{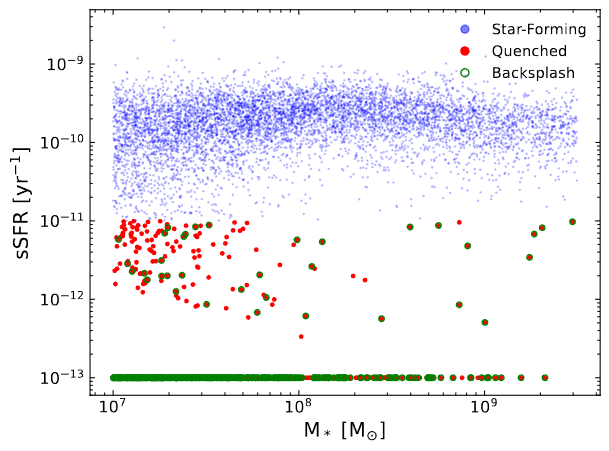}
    \caption{Specific star formation rate (sSFR) as a function of stellar mass for all simulated central galaxies in the TNG50 simulation, split into star-forming (blue filled circles) and quenched (red filled circles) galaxies, at $\mathrm{sSFR} = 10^{-11}~\mathrm{yr}^{-1}$. We also highlight the sample of backsplash galaxies (green open circles). Quenched galaxies with $\mathrm{SFR} \sim 0$ have been artificially placed at $\mathrm{sSFR} = 10^{-13}~\mathrm{yr}^{-1}$.
}
    \label{fig:ssfr_mass}
\end{figure}
\subsection{Sample Selection}
For this study, we employ the TNG50-1 simulation, which provides the highest mass resolution within the suite. We analyze the galaxy sample at $z=0$ from this simulation. 
The definition of field dwarf galaxy varies across different studies. For example,  Geha et al.\ (2012) defined field dwarf galaxies those located more than $1.5\,\mathrm{Mpc}$ from the nearest massive host, 
where a massive host refers to a galaxy with stellar mass greater than $2.5\times10^{10}\,M_{\odot}$. Here,  
following Benavides et al.\ (2025), we define field dwarf as any central galaxy with $10^{7} < M_{\star}/M_{\odot} < 10^{9.5}$ at $z=0$, and further exclude systems that have any satellite more massive than the host galaxy itself 
within 1.5\,Mpc. We consider this criterion to be sufficiently strict, since our definition of ``massive'' is relative to the host-galaxy stellar mass and our analysis focuses on low-mass galaxies.
To ensure adequate resolution, we impose a minimum stellar mass of $M_\star = 10^7\,M_\odot$, which guarantees that each galaxy is resolved with at least $\sim 100$ stellar particles \citep{2025OJAp....8E..43B}. For the upper limit, we adopt a stellar mass comparable to that of the Large Magellanic Cloud, corresponding to $\sim 10^{9.5}\,M_\odot$ \citep{2002AJ....124.2639V}. As a result, our study focuses on galaxies with stellar masses in the range $10^7 < M_\star/M_\odot < 10^{9.5}$.

At $z=0$, we classify galaxies according to their specific star formation rate (sSFR). Star-forming galaxies are defined as those with $\log\,(\mathrm{sSFR}/\mathrm{yr}^{-1}) \geq -10.5$, i.e., no more than 0.5 dex below the star-forming main sequence. Quenched galaxies are those with $\log\,(\mathrm{sSFR}/\mathrm{yr}^{-1}) \leq -11$, more than 1 dex below the ridge, as in \cite{lu2021hot}. Galaxies with intermediate values $-11 < \log\,(\mathrm{sSFR}/\mathrm{yr}^{-1}) < -10.5$ are classified as green-valley systems, representing a transitional population; for simplicity, we group green-valley galaxies together with the star-forming category throughout this work.

Figure~\ref{fig:ssfr_mass} shows the sSFR as a function of stellar mass for isolated dwarf galaxies at $z=0$. Following the above convention, star-forming and green-valley galaxies are shown in blue with $\log\,(\mathrm{sSFR}/\mathrm{yr}^{-1}) > -11$, while quenched galaxies are shown in red with $\log\,(\mathrm{sSFR}/\mathrm{yr}^{-1}) \leq -11$.

\begin{figure}[htbp]
    \centering
    \includegraphics[width=1.0\linewidth]{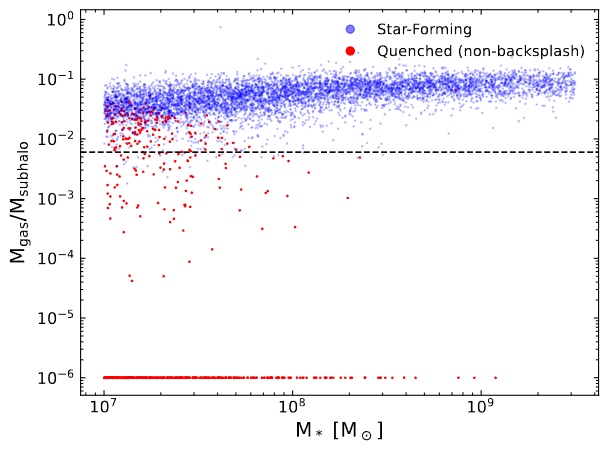}
    \caption{Distribution of the ratio of total gas mass to total subhalo mass as a function of galaxy stellar mass. Blue dots represent star-forming galaxies, while red dots represent quenched galaxies after excluding backsplash candidates. It can be seen that in the low-mass regime, some quenched galaxies still exhibit gas fractions comparable to those of star-forming galaxies. The horizontal dashed line at $\mathrm{M_{{gas}}/M_{{subhalo}} = 6\times10^{-3}}$ is shown in the figure; quenched galaxies above this threshold are classified as high-gas-fraction (HGF) galaxies, while those below it are classified as low-gas-fraction (LGF) galaxies. Galaxies with zero gas fractions are assigned a floor value of $\mathrm{M_{gas}/M_{subhalo}}=10^{-6}$.
}
    \label{fig:gas_ratio}
\end{figure}

\section{Analysis}
\subsection{Environmental Effects}

Although the galaxies in our sample are isolated at $z=0$, they are not necessarily isolated throughout their entire histories, and some may have been quenched by environmental processes. A subset may have previously been satellites of more massive host galaxies and later been ejected beyond the host halo as ``backsplash'' systems \citep[e.g.,][]{mamon2004origin,2005MNRAS.356.1327G, 2011MNRAS.412..529K}. In addition, even without being satellites, low-mass dwarfs can lose gas while traversing dense filaments and nodes of the cosmic web (``cosmic-web stripping''), which can also suppress star formation \citep[e.g.,][]{2013ApJ...763L..41B,10.1093/mnras/stac3282,benavides2025environmental}.
 
\begin{figure}[htbp]
  \centering
  \includegraphics[width=0.90\linewidth]{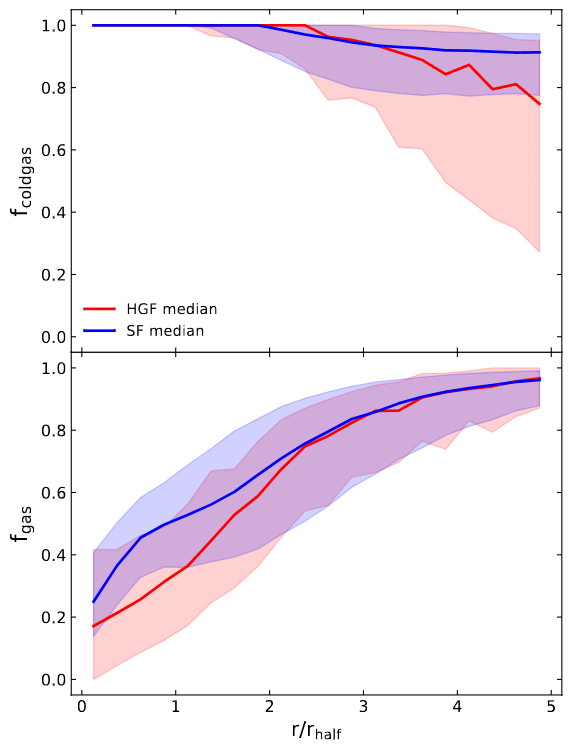}
  \caption{Radial gas distribution profiles for the HGF and SF samples within $5\,r_{\mathrm{half}}$. The top panel shows the cold-gas fraction, $M_{\mathrm{coldgas}}/M_{\mathrm{gas}}$, and the bottom panel shows the cumulative gas fraction relative to the total baryonic mass, $M_{\mathrm{gas}}/(M_{\mathrm{gas}}+M_\star)$, both as a function of radius in units of $r_{\mathrm{half}}$. The red curves represent the HGF sample and the blue curves represent the SF sample. In each panel, the solid curves denote the median profiles, while the shaded regions indicate the 16th--84th percentile ranges.}
  \label{gas_distribution_median}
\end{figure}

\begin{figure}[!tbp] 
  \centering
  \includegraphics[width=\columnwidth]{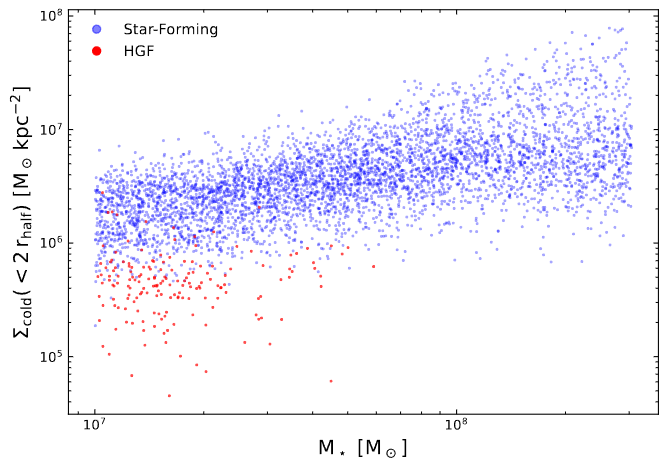}
  \caption{The distribution of the mean cold-gas surface density within $2\,r_{\rm half}$, as a function of stellar mass $M_\ast$ for low-mass star-forming galaxies (blue) and HGF quenched galaxies (red). While the two populations substantially overlap in total gas fraction (Fig.~\ref{fig:gas_ratio}), HGF systems exhibit systematically lower $\Sigma_{\rm cold}(<2\,r_{\rm half})$, implying a reduced central concentration of cold gas relative to star-forming galaxies at fixed $M_{\ast}$. 
  }
  \label{sf_hgf_surface_density}
\end{figure}

To minimize environmental influence in our quenching analysis, we further refine the selected sample based on merger-tree information. For each isolated quenched dwarf in TNG50, we trace the evolution of its specific star formation rate (sSFR) and FoF-central status along the main progenitor branch. We define the primary quenching interval as the period of steepest logarithmic decline in sSFR, identified using sliding windows of multiple lengths. This interval must end in the quenched regime, where $\mathrm{sSFR} \leq 10^{-11}\,\mathrm{yr}^{-1}$. 
Next, we compute the satellite ratio, $f_{\mathrm{sat}}$, defined as the fraction of snapshots within this interval during which the galaxy is not the FoF central. A $z=0$ isolated galaxy is classified as a backsplash candidate if $f_{\mathrm{sat}} \geq 40\%$. These backsplash candidates are highlighted with green circles in Figure~\ref{fig:ssfr_mass}. Consistent with \citet{benavides2025environmental}, we find that the more massive dwarfs in our sample are preferentially associated with this environmental quenching channel.

For the remaining quenched dwarfs, we use their gas content to assess whether gas removal is likely responsible. Figure~\ref{fig:gas_ratio} shows that, besides a gas-poor population, there exists a population of low-mass quenched galaxies that retain gas fractions comparable to those of star-forming dwarfs. These gas-rich, non-backsplash quenched dwarfs therefore constitute our primary targets for investigating quenching mechanisms that operate with minimal direct environmental gas removal.

\subsection{Mechanisms driving the quenching of low-mass, gas-rich galaxies. }
Gas-poor dwarf galaxies are readily quenched, whereas the quenching of gas-rich dwarfs is not yet understood. We therefore focus on mechanisms that can suppress star formation despite substantial remaining gas reservoirs. One possible explanation is that star formation is regulated by the dynamical state of the gas. 

\begin{figure*}[htbp]
  \centering

  \begin{subfigure}[b]{0.49\textwidth}
    \centering
    \includegraphics[width=\linewidth]{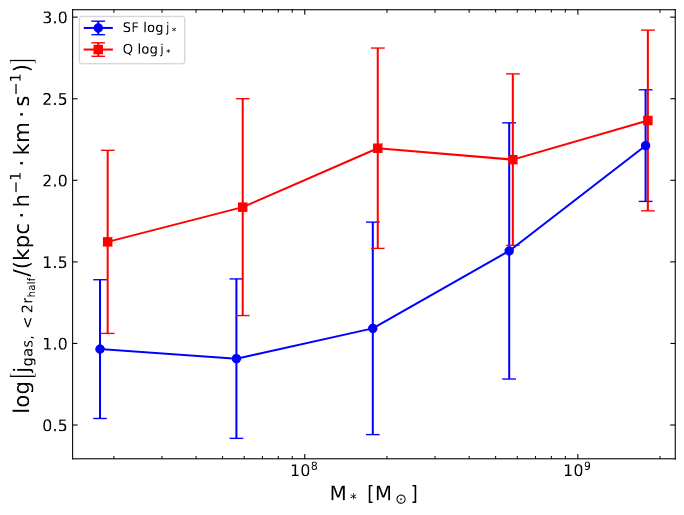}
    
  \end{subfigure}
  \begin{subfigure}[b]{0.49\textwidth}
    \centering
    \includegraphics[width=\linewidth]{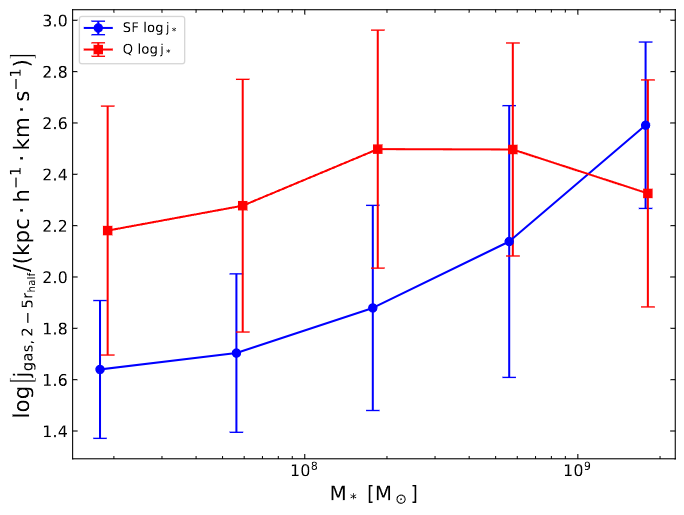}
  \end{subfigure}

  \vspace{0.4cm} 

  \begin{subfigure}[b]{0.49\textwidth}
    \centering
    \includegraphics[width=\linewidth]{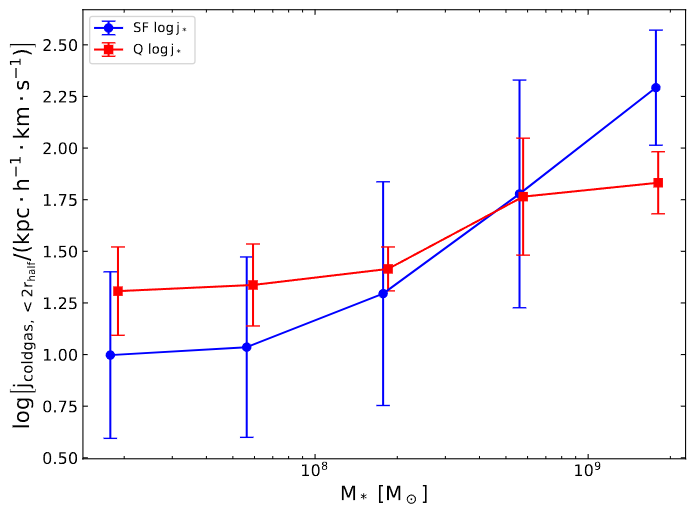}
  \end{subfigure}
  \begin{subfigure}[b]{0.49\textwidth}
    \centering
    \includegraphics[width=\linewidth]{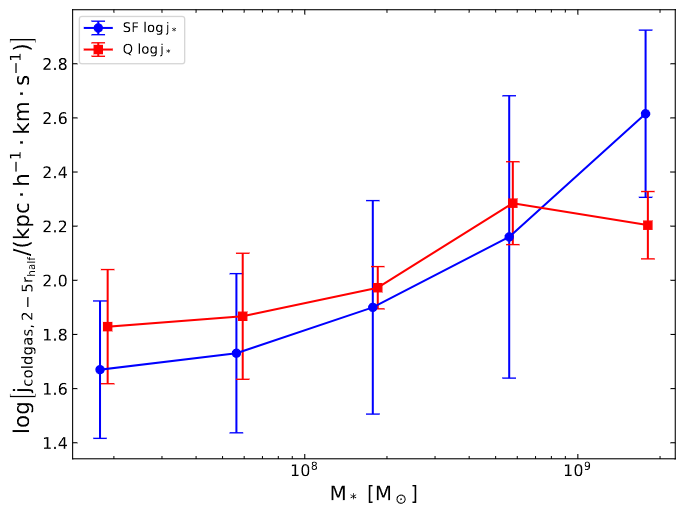}
  \end{subfigure}
  \caption{
The distribution of the average total gas specific angular momentum as a function of host galaxy stellar mass is shown, with quenched galaxies represented by the red line and star-forming galaxies by the blue line. 
The top-left panel plots the total gas specific angular momentum within twice the half-mass radius against the host stellar mass. 
The top-right panel plots the total gas specific angular momentum in the radial range between two and five times the half-mass radius against the host stellar mass. 
The bottom-left panel represents the results for the specific angular momentum of the cold gas within twice the half-mass radius, while the bottom-right panel is the results for the cold gas specific angular momentum between two and five times the half-mass radius.
}
  \label{fig:2x2grid}
\end{figure*}

We find that a substantial subset of these low-mass quenched dwarfs remain gas rich, with total gas fractions overlapping those of star-forming galaxies of similar stellar mass. Motivated by this diversity in gas content, we subdivide the quenched population according to their gas fraction into high--gas-fraction (HGF, with gas fraction $\mathrm{M_{gas}/M_{subhalo}} > 6 \times 10^{-3}$, the samples above the horizontal dashed line in Fig.~\ref{fig:gas_ratio}) and low--gas-fraction (LGF, with gas fraction $\mathrm{M_{gas}/M_{subhalo}} \leq 6 \times 10^{-3}$) systems. For convenience, we restrict our main analysis to galaxies with $M_{\ast} \leq 10^{8}\,M_{\odot}$, which we hereafter refer to as low-mass galaxies.We examined the spatial distribution of gas in these gas-rich quenched systems at $z=0$, in comparison with star-forming galaxies of comparable mass.

In Fig.~\ref{gas_distribution_median}, we show the median radial profiles of the star-forming (SF) and HGF galaxies, with shaded regions indicating the 16th--84th percentile ranges, and present (i) the radial variation of the cold-gas fraction, $\mathrm{M_{coldgas}/M_{gas}}$, and (ii) the cumulative gas fraction relative to the total baryonic mass, $\mathrm{M_{gas}/(M_{gas}+M_\star)}$, both as a function of radius in units of $r_{\mathrm{half}}$.

Here we provide the definition of the gas temperature used in TNG:

\begin{align}
x_e &\equiv \frac{n_e}{n_{\mathrm{H}}},\\[2pt]
\mu &= \frac{4}{1 + 3X_{\mathrm{H}} + 4X_{\mathrm{H}}x_e}\cdot m_{\mathrm{p}},\\[2pt]
T &= (\gamma - 1)\cdot \frac{u}{k_B}\cdot \frac{\texttt{UnitEnergy}}{\texttt{UnitMass}}\cdot \mu .
\end{align}

Here $u$ is the gas internal energy per unit mass and $x_e$ is the electron abundance (from the TNG snapshot fields \texttt{InternalEnergy} and \texttt{ElectronAbundance}, respectively). In TNG, $\gamma=5/3$ and $X_{\mathrm{H}}=0.76$, where $\gamma$ is the adiabatic index and $X_{\mathrm{H}}$ is the hydrogen mass fraction. The factor $\texttt{UnitEnergy}/\texttt{UnitMass}$ converts $u$ from code units to CGS units. For the TNG default units,

\begin{equation}
\begin{aligned}
\frac{\texttt{UnitEnergy}}{\texttt{UnitMass}}
&= \texttt{UnitVelocity}^2 \\
&= \left(1\,\mathrm{km\cdot s^{-1}}\right)^2 \\
&= 10^{10}\,\mathrm{cm^2\cdot s^{-2}}
= 10^{10}\,\mathrm{erg\,g^{-1}} .
\end{aligned}
\end{equation}

Accordingly, we define ``cold gas'' as gas that either has a temperature $T < 2\times 10^{4}\,\mathrm{K}$ or is star-forming. In \textsc{TNG}, each gas cell is assigned a $\mathrm{StarFormationRate}$; we classify gas cells with $\mathrm{StarFormationRate} > 0$ as star-forming
\citep{10.1093/mnras/stab3167,wang2024larger,Wright2024}.

 As shown in Fig.~\ref{gas_distribution_median}, although the gas within $2\,r_{\mathrm{half}}$ in HGF galaxies is almost entirely in the cold phase, their total gas content in this inner region is lower than that of star-forming galaxies. At larger radii ($2$--$5\,r_{\mathrm{half}}$), HGF galaxies contain a total gas mass comparable to that of star-forming systems, but with a markedly lower cold-gas fraction. This suggests that, compared with star-forming galaxies, HGF systems contain less gas overall that is available for star formation. We further quantify the inner cold-gas distribution by comparing the cold-gas surface density profiles of low-mass star-forming galaxies to those of HGF systems. Figure~\ref{sf_hgf_surface_density} shows the distribution of the mean cold-gas surface density within $2\,r_{\mathrm{half}}$ as a function of $M_\ast$. Although HGF galaxies substantially overlap with low-mass star-forming galaxies in total gas fraction (Fig.~\ref{fig:gas_ratio}), they exhibit systematically lower cold-gas surface densities within $2\,r_{\mathrm{half}}$. This suggests that their gas is preferentially heated and redistributed to larger radii, leaving a diluted cold component in the inner regions and thereby helping explain why their sizable gas reservoirs do not translate into sustained star formation. This naturally raises the question of what drives such a configuration. We therefore turn to mechanisms that can suppress star formation despite the presence of substantial remaining gas reservoirs. One possible explanation is that star formation is regulated by the dynamical state of the gas. In particular, excessively high specific angular momentum can stabilize the gas against collapse and reduce the efficiency of inflow, cooling, and fragmentation, thereby inhibiting star formation. A similar connection between quiescence and high angular momentum has also been discussed for massive quenched galaxies \citep{2022MNRAS.509.2707L}. We measure the specific angular momentum of both the total gas and the cold-gas component within five stellar half-mass radii (See Figure~\ref{fig:2x2grid}). As shown in the figure, at the low-mass end, quenched galaxies (red) consistently exhibit higher gas specific angular momentum than star-forming galaxies (blue), regardless of whether the total or cold-gas component is considered. 
 The physical origin of this excess remains unclear, motivating further investigation into the underlying mechanisms, starting with whether internal feedback processes contribute to the quenching of these galaxies.
 
Motivated by our hypothesis we statistically quantify gas flows in the SF, HGF, and LGF systems using the shell-based gas flow rate estimator at $r$ from \citet{Wright2024},
\begin{equation}
\dot{M}_k(r)=\frac{1}{\Delta r}\sum_{i\in k}\left(m_i\,\frac{\vec{v}_{ij}\cdot\vec{r}_{ij}}{|\vec{r}_{ij}|}\right),
\end{equation}

\begin{figure*}[!tbp]
  \centering  \includegraphics[width=0.90\linewidth,height=0.40\textheight,keepaspectratio]{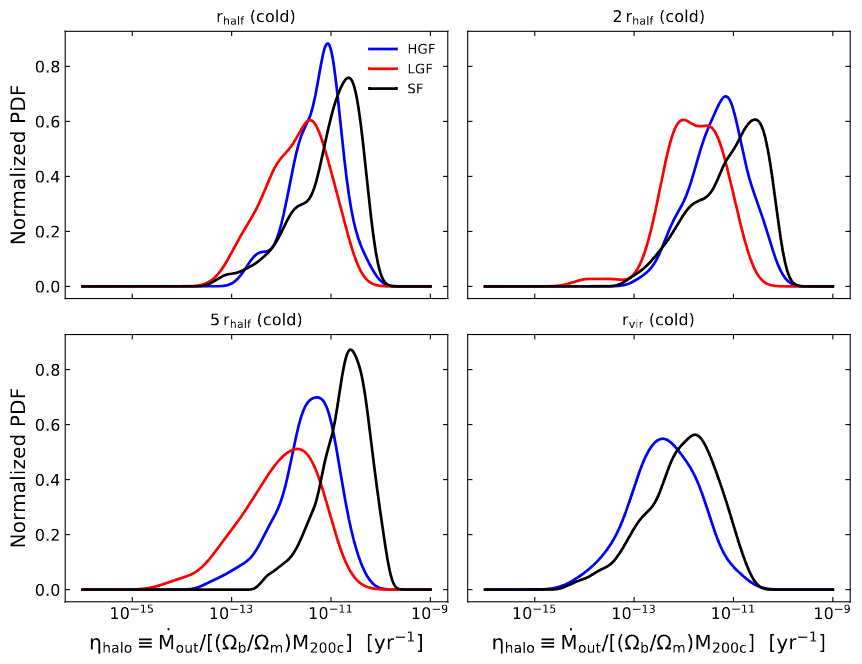}
  \caption{Probability distribution functions (PDFs) of the cold gas outflow efficiency ($\eta_{\text{halo}}$) for star-forming (black lines), high gas fraction (HGF; blue lines) and low gas fraction (LGF; orange lines) galaxies. The outflow efficiency is normalized by theoretical total baryon mass, defined as $(\Omega_{\mathrm{b}} / \Omega_{\mathrm {m}}) {M}_{200c}$, where we take the virial mass to be $M_{200c}$, defined as the total mass enclosed within a sphere whose mean density is 200 times the critical density of the Universe at the halo redshift (i.e., $\mathrm{Group\_M\_Crit200}$ in the IllustrisTNG group catalogue).
 Each panel displays the distribution measured at a different radius: $1\,r_{\rm half}$ (top-left), $2\,r_{\rm half}$ (top-right), $5\,r_{\rm half}$ (bottom-left), and the virial radius ($r_{\text{vir}}$; bottom-right). A curve is shown only when at least three galaxies in that radial bin have a positive outflow efficiency. At the virial radius, fewer than three LGF galaxies exhibit a measurable outflow, and therefore no LGF curve is shown in that panel.}
  \label{outflow_rate}
\end{figure*}

here we adopt a fractional shell thickness of $\Delta r = 0.2\,r$ (i.e. $r\pm 0.1r$). The index $j$ denotes the subhalo center, and $m_i$ is the mass of gas cell $i$. Here $\vec{r}_{ij}\equiv \vec{r}_i-\vec{r}_j$ and $\vec{v}_{ij}\equiv \vec{v}_i-\vec{v}_j$,  $(\vec{r}_j,\vec{v}_j)$ are taken to be the halo-center position and bulk velocity from the group catalog and $(\vec{r}_i,\vec{v}_i)$ is the gas cell's position and velocity. We classify cells with $\dot{M}_k>0$ as \textit{outflow}($\dot{M}_{out}$) and $\dot{M}_k<0$ as \textit{inflow}. Our statistical analysis of cold-gas outflow velocities reveals significant differences among SF, HGF and LGF galaxies. Specifically, within $5\,r_{\mathrm{half}}$, the outflow rates of the SF, HGF and LGF galaxies differ systematically, with pairwise offsets of roughly $0.5$~dex, as shown in the Figure~\ref{outflow_rate}. Moreover, while cold gas is effectively absent in LGF galaxies near the virial radius ($R_{\mathrm{vir}}$), SF and HGF galaxies continue to display measurable cold-gas outflows at this distance. 
Notably, in the TNG model, outflows are launched directly from star-forming gas through the injection of stellar wind energy\citep{Pillepich2018}, which naturally helps explain the particularly high outflow rates seen in SF galaxies. However, among HGF and LGF galaxies with comparable sSFRs, HGF systems still exhibit systematically higher outflow rates than LGF systems. This suggests that stellar feedback may also play an important role in driving gas outflows in HGF galaxies.

\begin{figure}[htbp]
  \centering
  \includegraphics[width=0.90\linewidth]{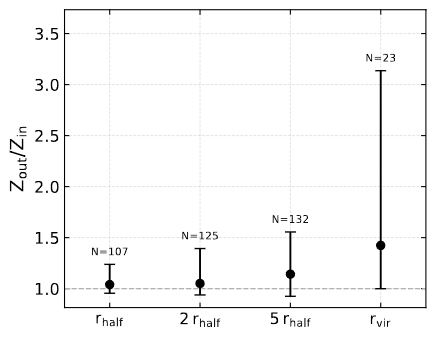}
  \caption{Ratio of the metallicity of outflowing to inflowing cold gas as a function of radius for HGF quenched dwarf galaxies. The four radial bins correspond to
$1\,r_{\mathrm{half}}$, $2\,r_{\mathrm{half}}$, $5\,r_{\mathrm{half}}$,
and $r_{\mathrm{vir}}$. Black points show the median $Z_{\mathrm{out}}/Z_{\mathrm{in}}$
for galaxies in which both inflows and outflows are present at the
corresponding radius, and for which both the inflowing and outflowing metallicities are estimated from at least five gas cells; $N$ denotes the number of galaxies satisfying these criteria in each radial bin. The vertical error bars indicate the 16th–84th percentile range. The dashed horizontal line marks $Z_{\mathrm{out}}/Z_{\mathrm{in}}=1$.
At all radii the median ratio exceeds unity, indicating that the outflowing gas is systematically more metal-enriched than the inflowing component,
consistent with stellar-feedback–driven winds redistributing enriched ISM gas to large radii.}
  \label{Metallicity}
\end{figure}

To test whether the distinct gas properties in HGF dwarfs are indeed driven by stellar feedback, we compare the metallicities of outflowing and inflowing cold gas \citep{Pillepich2018} at the four characteristic radii discussed above ($1\,r_{\mathrm{half}}$, $2\,r_{\mathrm{half}}$, $5\,r_{\mathrm{half}}$, and $r_{\mathrm{vir}}$) shown in Figure~\ref{Metallicity}. For galaxies in which both inflows and outflows are present, we find that the outflowing gas is systematically more metal-enriched than the inflowing component at all radii. This systematic metallicity excess strongly suggests that the outflows are predominantly composed of gas that has been processed in the star-forming ISM and enriched by previous generations of stars, providing further evidence that the extended cold-gas outflows in HGF quenched dwarfs are primarily driven by stellar feedback.

\begin{figure}[tbp!]
  \centering
  \includegraphics[width=0.9\linewidth]{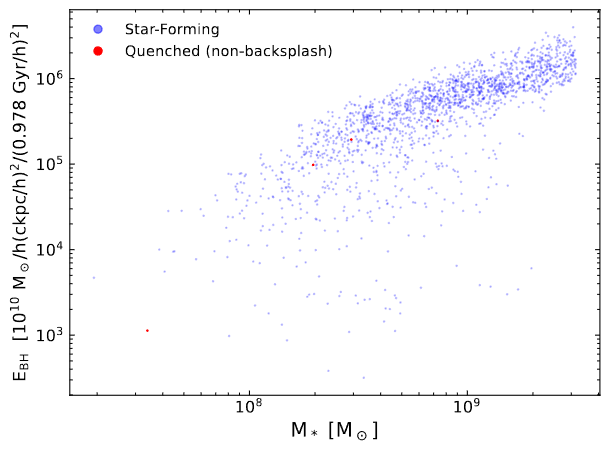}
  \caption{Distribution of galaxy stellar mass versus black hole energy injection rate. Blue dots represent star-forming galaxies, while red dots represent quenched galaxies after excluding backsplash candidates. AGN are rare in these quenched systems; when present, their energy injection rates are comparable to those of star-forming galaxies at similar stellar masses.}
  \label{AGN}
\end{figure}

\begin{figure}[htbp]
  \centering
  \includegraphics[width=1.0\linewidth]{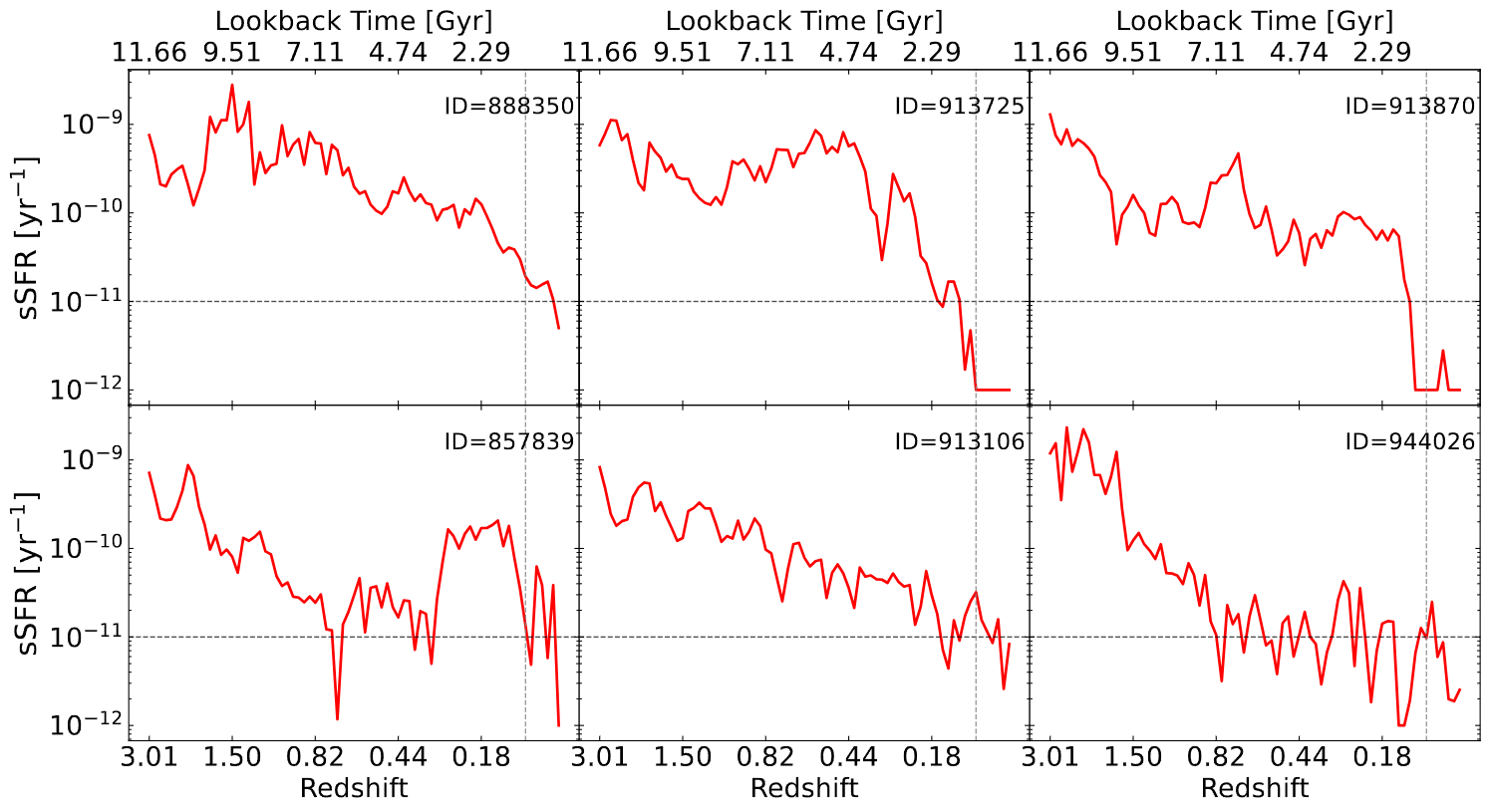}
  \caption{Star-formation histories of six HGF galaxies shown as the evolution of sSFR with redshift. In each panel, the upper horizontal axis indicates the lookback time, while the lower horizontal axis shows the redshift. The vertical axis gives the sSFR; the dashed horizontal line marks the quenching threshold adopted in this work, and the dashed vertical line indicates the epoch corresponding to a lookback time of $1\,\mathrm{Gyr}$. The upper three panels show representative HGF galaxies that are permanently quenched, or at least show no clear evidence of temporary quenching, which constitute the majority of the HGF population, whereas the lower three panels show systems that are likely only temporarily quenched. All six galaxies are nevertheless in a quenched state at $z=0$.}
  \label{6SFH_sample}
\end{figure}

Having established that the outflowing gas in HGF dwarfs is more metal-enriched
than the inflowing component, this suggests a close connection between the outflows and stellar feedback, since feedback-driven winds are expected to preferentially expel chemically enriched gas from star-forming regions. To further test whether the outflows in HGF dwarfs are primarily driven by stellar feedback, we also assess the potential role of AGN feedback by measuring the cumulative
AGN energy injection for our sample (Fig.~\ref{AGN}). The statistics show that
AGN activity is extremely rare in the low-mass regime considered here: the
energy input is nearly zero for the vast majority of dwarf galaxies, and we
find no significant difference between quenched and star-forming systems.
This is consistent with the IllustrisTNG black hole model \citep{weinberger2018supermassive}. Motivated by this, we do not pursue a detailed investigation of AGN-related quenching mechanisms in this work.

We further examined the sSFR evolution of individual HGF dwarfs to test whether their quenching is temporary or long-lived. If these galaxies were only temporarily quenched, one would expect bursty star-formation histories and subsequent rejuvenation episodes. We find that most HGF galaxies behave like the upper three examples shown in Fig.~\ref{6SFH_sample}: their sSFR declines gradually with time. Among the 173 HGF galaxies, 59 have remained quenched for longer than $1\,\mathrm{Gyr}$, as illustrated by the two galaxies on the right-hand side of the top row. Only a small minority of systems appear to be merely temporarily quenched at the present epoch (see the lower three examples in Fig.~\ref{6SFH_sample}), accounting for about 12 out of our 173 HGF galaxies. Since our primary focus is on the galaxy population at $z=0$, this result supports the conclusion that the majority of HGF galaxies in our sample are permanently quenched.

In addition, we quantified the incidence of more massive neighboring galaxies in a relatively distant shell around each central galaxy (1.5--2.0 cMpc). Since our isolation criterion is defined within 1.5 cMpc, this larger-scale search provides an additional test of their degree of isolation. Specifically, we counted neighbors with stellar mass exceeding that of the central within this radial range and converted the counts into a shell number density. The results are summarised in Table~\ref{Mpc}. We find that a substantially larger fraction of HGF centrals have no more massive neighbors in this shell compared to the LGF sample, indicating that HGF galaxies tend to inhabit more isolated environments. The fact that those permanently quenched HGF systems exist raises an important question: whether stellar feedback alone is sufficient to quench them, or whether additional environmental effects are still at work.

\begin{table}
\centering
\caption{Fraction of central galaxies with at least one more-massive neighbor in the 1.5--2.0 cMpc shell at $z=0$. Column~(1) lists the sample; column~(2) gives the total number of centrals, $N_{\rm gal}$; column~(3) gives the number of centrals with at least one more-massive neighbor within this shell, $N(\rho>0)$; column~(4) lists the corresponding fraction, $f_{>0}\equiv N(\rho>0)/N_{\rm gal}$; and column~(5) gives the complementary fraction with no such neighbors, $f_{0}\equiv 1-f_{>0}$.}
\label{Mpc}
\begin{tabular}{lcccc}
\hline
Sample & $N_{\rm gal}$ & $N(\rho>0)$ & $f_{>0}$ & $f_0$ \\
\hline
HGF & 173 & 34  & 0.197 & 0.803 \\
LGF & 480 & 174 & 0.363 & 0.637 \\
\hline
\end{tabular}
\end{table}


\section{Conclusions}
In this work, we use the TNG50 simulation at $z=0$ to study isolated central dwarf galaxies and to investigate the origin of quenching in the absence of strong environmental processing. Throughout, we classify our galaxies into star-forming systems and two quenched sub-populations, namely high--gas-fraction (HGF) and low--gas-fraction (LGF) quenched dwarfs. Our main findings are:

\begin{enumerate}[label=(\roman*)]
    \item By examining the gas content of isolated quenched dwarfs as a
    function of stellar mass, we find that, while most quenched systems
    are gas-poor, a non-negligible fraction still retain gas fractions
    comparable to those of star-forming dwarfs. This population of
    gas-rich quenched dwarfs is more naturally explained by internal
    quenching processes rather than by environmentally driven gas
    removal.

    \item Motivated by the overlap in gas fractions between a subset of
    quenched dwarfs and star-forming galaxies, we divide the quenched
    population into HGF and LGF
    systems. A further analysis of the gas distribution in HGF galaxies
    shows that most of their gas is heated and preferentially resides at
    large radii, rather than in the central star-forming region (Fig.~\ref{gas_distribution_median}). A direct
    comparison of the cold-gas surface density within $2\,r_{\rm half}$
    between HGF and star-forming galaxies (Fig.~\ref{sf_hgf_surface_density}) further supports this picture:
    HGF systems exhibit systematically lower inner cold-gas surface
    densities, consistent with heating and redistribution of cold gas
    by stellar feedback.

    \item We find that, at the low-mass end, quenched dwarfs exhibit
    systematically higher gas specific angular momentum than
    star-forming galaxies of the same stellar mass (Fig.~\ref{fig:2x2grid}). This excess angular
    momentum implies that the gas reservoir in quenched systems is, on
    average, more extended and less efficiently drained of angular
    momentum, consistent with a reduced ability to build up a dense,
    star-forming central component.

    \item Consistent with a stellar-feedback-driven scenario, we find that
    HGF galaxies show more disturbed gas kinematics than LGF systems,
    including substantially higher outflow velocities (Fig.~\ref{outflow_rate}). Moreover,
    the outflowing gas is systematically more metal-rich than the
    inflowing component (Fig.~\ref{Metallicity}) wherever both are present, indicating that the
    outflows contain ISM-processed material that has been expelled from
    the star-forming regions by stellar feedback.

    \item We discuss the possible role of AGN in the text; however, given the specific implementation of the AGN model in IllustrisTNG, our analysis does not allow a clean isolation or quantification of AGN-driven effects in these low-mass systems. This does not imply that AGN are irrelevant: AGN feedback could still influence a minority of dwarf galaxies, and its impact in the low-mass regime remains an open question that warrants further investigation.
    
    \item We investigated the sSFR histories of HGF galaxies in order to assess whether their quenching is temporary or long-lived. Although a small subset of HGF galaxies appear to show some tendency to return to the star-forming population after quenching, they still remain quenched at $z=0$, and we cannot determine whether they will definitely resume star formation in the future. The majority, however, remain quenched, with evolutionary histories more consistent with long-lived, and likely permanent, quenching.
\end{enumerate}

\begin{acknowledgements}
We thank the referee for comments and suggestions that improved the paper. We also thank Qi Guo and Dandan Xu for helpful comments and discussions. 
This work is supported by the National SKA Program of China (Nos. 2022SKA0110100 and 2022SKA0110101), the National Natural Science Foundation of China (NSFC) International (Regional) Cooperation and Exchange Project (No. 12361141814), the Specialized Research Fund for State Key Laboratory of Radio Astronomy and Technology, and the National Astronomical Observatories, Chinese Academy of Science (No. E5ZB0901). SL acknowledges support by the NSFC grant (no. 12588202, 12473015). 

\end{acknowledgements}

%
%

\bibliographystyle{raa}
\bibliography{bibtex}

@ARTICLE{2017MNRAS.470..283W,
       author = {{Wright}, A.~H. and {Robotham}, A.~S.~G. and {Driver}, S.~P. and {Alpaslan}, M. and {Andrews}, S.~K. and {Baldry}, I.~K. and {Bland-Hawthorn}, J. and {Brough}, S. and {Brown}, M.~J.~I. and {Colless}, M. and {da Cunha}, E. and {Davies}, L.~J.~M. and {Graham}, Alister W. and {Holwerda}, B.~W. and {Hopkins}, A.~M. and {Kafle}, P.~R. and {Kelvin}, L.~S. and {Loveday}, J. and {Maddox}, S.~J. and {Meyer}, M.~J. and {Moffett}, A.~J. and {Norberg}, P. and {Phillipps}, S. and {Rowlands}, K. and {Taylor}, E.~N. and {Wang}, L. and {Wilkins}, S.~M.},
        title = "{Galaxy And Mass Assembly (GAMA): the galaxy stellar mass function to z = 0.1 from the r-band selected equatorial regions}",
      journal = {\mnras},
         year = 2017,
        month = sep,
       volume = {470},
       number = {1},
        pages = {283-302},
          doi = {10.1093/mnras/stx1149},
archivePrefix = {arXiv},
       eprint = {1705.04074},
 primaryClass = {astro-ph.GA},
       adsurl = {https://ui.adsabs.harvard.edu/abs/2017MNRAS.470..283W}
}

@ARTICLE{2025ApJ...994..231K,
       author = {{Kado-Fong}, Erin and {Mao}, Yao-Yuan and {Asali}, Yasmeen and {Geha}, Marla and {Wechsler}, Risa H. and {de los Reyes}, Mithi A.~C. and {Wang}, Yunchong and {Nadler}, Ethan O. and {Kallivayalil}, Nitya and {Tollerud}, Erik J. and {Weiner}, Benjamin},
        title = "{SAGAbg. III. Environmental Stellar Mass Functions, Self-quenching, and the Stellar-to-halo Mass Relation in the Dwarf Galaxy Regime}",
      journal = {\apj},
         year = 2025,
        month = dec,
       volume = {994},
       number = {2},
          eid = {231},
        pages = {231},
          doi = {10.3847/1538-4357/ae102d},
archivePrefix = {arXiv},
       eprint = {2509.20444},
 primaryClass = {astro-ph.GA},
       adsurl = {https://ui.adsabs.harvard.edu/abs/2025ApJ...994..231K}
}

@ARTICLE{Zheng2025,
       author = {{Zheng}, Haonan and {Jiang}, Fangzhou and {Liao}, Shihong and {Libeskind}, Noam I.},
        title = "{HIDES -- I. The population and diversity of HI-rich 'dark' galaxies in the Hestia and Auriga simulations}",
      journal = {arXiv e-prints},
         year = 2025,
        month = nov,
          eid = {arXiv:2511.16726},
        pages = {arXiv:2511.16726},
          doi = {10.48550/arXiv.2511.16726},
archivePrefix = {arXiv},
       eprint = {2511.16726},
 primaryClass = {astro-ph.GA},
       adsurl = {https://ui.adsabs.harvard.edu/abs/2025arXiv251116726Z}
}

@ARTICLE{2018MNRAS.478..548S,
       author = {{Simpson}, Christine M. and {Grand}, Robert J.~J. and {G{\'o}mez}, Facundo A. and {Marinacci}, Federico and {Pakmor}, R{\"u}diger and {Springel}, Volker and {Campbell}, David J.~R. and {Frenk}, Carlos S.},
        title = "{Quenching and ram pressure stripping of simulated Milky Way satellite galaxies}",
      journal = {\mnras},
         year = 2018,
        month = jul,
       volume = {478},
       number = {1},
        pages = {548-567},
          doi = {10.1093/mnras/sty774},
archivePrefix = {arXiv},
       eprint = {1705.03018},
 primaryClass = {astro-ph.GA},
       adsurl = {https://ui.adsabs.harvard.edu/abs/2018MNRAS.478..548S}
}

@ARTICLE{2013ApJ...763L..41B,
       author = {{Ben{\'\i}tez-Llambay}, Alejandro and {Navarro}, Julio F. and {Abadi}, Mario G. and {Gottl{\"o}ber}, Stefan and {Yepes}, Gustavo and {Hoffman}, Yehuda and {Steinmetz}, Matthias},
        title = "{Dwarf Galaxies and the Cosmic Web}",
      journal = {\apjl},
         year = 2013,
        month = feb,
       volume = {763},
       number = {2},
          eid = {L41},
        pages = {L41},
          doi = {10.1088/2041-8205/763/2/L41},
archivePrefix = {arXiv},
       eprint = {1211.0536},
 primaryClass = {astro-ph.CO},
       adsurl = {https://ui.adsabs.harvard.edu/abs/2013ApJ...763L..41B}
}

@ARTICLE{2005MNRAS.356.1327G,
       author = {{Gill}, Stuart P.~D. and {Knebe}, Alexander and {Gibson}, Brad K.},
        title = "{The evolution of substructure - III. The outskirts of clusters}",
      journal = {\mnras},
         year = 2005,
        month = feb,
       volume = {356},
       number = {4},
        pages = {1327-1332},
          doi = {10.1111/j.1365-2966.2004.08562.x},
archivePrefix = {arXiv},
       eprint = {astro-ph/0404427},
 primaryClass = {astro-ph},
       adsurl = {https://ui.adsabs.harvard.edu/abs/2005MNRAS.356.1327G}
}

@ARTICLE{2011MNRAS.412..529K,
       author = {{Knebe}, Alexander and {Libeskind}, Noam I. and {Knollmann}, Steffen R. and {Martinez-Vaquero}, Luis A. and {Yepes}, Gustavo and {Gottl{\"o}ber}, Stefan and {Hoffman}, Yehuda},
        title = "{The luminosities of backsplash galaxies in constrained simulations of the Local Group}",
      journal = {\mnras},
         year = 2011,
        month = mar,
       volume = {412},
       number = {1},
        pages = {529-536},
          doi = {10.1111/j.1365-2966.2010.17924.x},
archivePrefix = {arXiv},
       eprint = {1010.5670},
 primaryClass = {astro-ph.CO},
       adsurl = {https://ui.adsabs.harvard.edu/abs/2011MNRAS.412..529K}
}

@ARTICLE{2019MNRAS.490.3234N,
       author = {{Nelson}, Dylan and {Pillepich}, Annalisa and {Springel}, Volker and {Pakmor}, R{\"u}diger and {Weinberger}, Rainer and {Genel}, Shy and {Torrey}, Paul and {Vogelsberger}, Mark and {Marinacci}, Federico and {Hernquist}, Lars},
        title = "{First results from the TNG50 simulation: galactic outflows driven by supernovae and black hole feedback}",
      journal = {\mnras},
         year = 2019,
        month = dec,
       volume = {490},
       number = {3},
        pages = {3234-3261},
          doi = {10.1093/mnras/stz2306},
archivePrefix = {arXiv},
       eprint = {1902.05554},
 primaryClass = {astro-ph.GA},
       adsurl = {https://ui.adsabs.harvard.edu/abs/2019MNRAS.490.3234N}
}

@ARTICLE{2010MNRAS.401..791S,
       author = {{Springel}, Volker},
        title = "{E pur si muove: Galilean-invariant cosmological hydrodynamical simulations on a moving mesh}",
      journal = {\mnras},
         year = 2010,
        month = jan,
       volume = {401},
       number = {2},
        pages = {791-851},
          doi = {10.1111/j.1365-2966.2009.15715.x},
archivePrefix = {arXiv},
       eprint = {0901.4107},
 primaryClass = {astro-ph.CO},
       adsurl = {https://ui.adsabs.harvard.edu/abs/2010MNRAS.401..791S}
}

@ARTICLE{2018MNRAS.475..648P,
       author = {{Pillepich}, Annalisa and {Nelson}, Dylan and {Hernquist}, Lars and {Springel}, Volker and {Pakmor}, R{\"u}diger and {Torrey}, Paul and {Weinberger}, Rainer and {Genel}, Shy and {Naiman}, Jill P. and {Marinacci}, Federico and {Vogelsberger}, Mark},
        title = "{First results from the IllustrisTNG simulations: the stellar mass content of groups and clusters of galaxies}",
      journal = {\mnras},
         year = 2018,
        month = mar,
       volume = {475},
       number = {1},
        pages = {648-675},
          doi = {10.1093/mnras/stx3112},
archivePrefix = {arXiv},
       eprint = {1707.03406},
 primaryClass = {astro-ph.GA},
       adsurl = {https://ui.adsabs.harvard.edu/abs/2018MNRAS.475..648P}
}

@ARTICLE{2018MNRAS.475..624N,
       author = {{Nelson}, Dylan and {Pillepich}, Annalisa and {Springel}, Volker and {Weinberger}, Rainer and {Hernquist}, Lars and {Pakmor}, R{\"u}diger and {Genel}, Shy and {Torrey}, Paul and {Vogelsberger}, Mark and {Kauffmann}, Guinevere and {Marinacci}, Federico and {Naiman}, Jill},
        title = "{First results from the IllustrisTNG simulations: the galaxy colour bimodality}",
      journal = {\mnras},
         year = 2018,
        month = mar,
       volume = {475},
       number = {1},
        pages = {624-647},
          doi = {10.1093/mnras/stx3040},
archivePrefix = {arXiv},
       eprint = {1707.03395},
 primaryClass = {astro-ph.GA},
       adsurl = {https://ui.adsabs.harvard.edu/abs/2018MNRAS.475..624N}
}

@ARTICLE{2018MNRAS.480.5113M,
       author = {{Marinacci}, Federico and {Vogelsberger}, Mark and {Pakmor}, R{\"u}diger and {Torrey}, Paul and {Springel}, Volker and {Hernquist}, Lars and {Nelson}, Dylan and {Weinberger}, Rainer and {Pillepich}, Annalisa and {Naiman}, Jill and {Genel}, Shy},
        title = "{First results from the IllustrisTNG simulations: radio haloes and magnetic fields}",
      journal = {\mnras},
         year = 2018,
        month = nov,
       volume = {480},
       number = {4},
        pages = {5113-5139},
          doi = {10.1093/mnras/sty2206},
archivePrefix = {arXiv},
       eprint = {1707.03396},
 primaryClass = {astro-ph.CO},
       adsurl = {https://ui.adsabs.harvard.edu/abs/2018MNRAS.480.5113M}
}

@ARTICLE{2018MNRAS.477.1206N,
       author = {{Naiman}, Jill P. and {Pillepich}, Annalisa and {Springel}, Volker and {Ramirez-Ruiz}, Enrico and {Torrey}, Paul and {Vogelsberger}, Mark and {Pakmor}, R{\"u}diger and {Nelson}, Dylan and {Marinacci}, Federico and {Hernquist}, Lars and {Weinberger}, Rainer and {Genel}, Shy},
        title = "{First results from the IllustrisTNG simulations: a tale of two elements - chemical evolution of magnesium and europium}",
      journal = {\mnras},
         year = 2018,
        month = jun,
       volume = {477},
       number = {1},
        pages = {1206-1224},
          doi = {10.1093/mnras/sty618},
archivePrefix = {arXiv},
       eprint = {1707.03401},
 primaryClass = {astro-ph.GA},
       adsurl = {https://ui.adsabs.harvard.edu/abs/2018MNRAS.477.1206N}
}

@ARTICLE{2018MNRAS.475..676S,
       author = {{Springel}, Volker and {Pakmor}, R{\"u}diger and {Pillepich}, Annalisa and {Weinberger}, Rainer and {Nelson}, Dylan and {Hernquist}, Lars and {Vogelsberger}, Mark and {Genel}, Shy and {Torrey}, Paul and {Marinacci}, Federico and {Naiman}, Jill},
        title = "{First results from the IllustrisTNG simulations: matter and galaxy clustering}",
      journal = {\mnras},
         year = 2018,
        month = mar,
       volume = {475},
       number = {1},
        pages = {676-698},
          doi = {10.1093/mnras/stx3304},
archivePrefix = {arXiv},
       eprint = {1707.03397},
 primaryClass = {astro-ph.GA},
       adsurl = {https://ui.adsabs.harvard.edu/abs/2018MNRAS.475..676S}
}

@ARTICLE{1985ApJ...292..371D,
       author = {{Davis}, M. and {Efstathiou}, G. and {Frenk}, C.~S. and {White}, S.~D.~M.},
        title = "{The evolution of large-scale structure in a universe dominated by cold dark matter}",
      journal = {\apj},
         year = 1985,
        month = may,
       volume = {292},
        pages = {371-394},
          doi = {10.1086/163168},
       adsurl = {https://ui.adsabs.harvard.edu/abs/1985ApJ...292..371D}
}

@ARTICLE{2001MNRAS.328..726S,
       author = {{Springel}, Volker and {White}, Simon D.~M. and {Tormen}, Giuseppe and {Kauffmann}, Guinevere},
        title = "{Populating a cluster of galaxies - I. Results at z=0}",
      journal = {\mnras},
         year = 2001,
        month = dec,
       volume = {328},
       number = {3},
        pages = {726-750},
          doi = {10.1046/j.1365-8711.2001.04912.x},
archivePrefix = {arXiv},
       eprint = {astro-ph/0012055},
 primaryClass = {astro-ph},
       adsurl = {https://ui.adsabs.harvard.edu/abs/2001MNRAS.328..726S}
}

@ARTICLE{Weinberger2017,
       author = {{Weinberger}, Rainer and {Springel}, Volker and {Hernquist}, Lars and {Pillepich}, Annalisa and {Marinacci}, Federico and {Pakmor}, R{\"u}diger and {Nelson}, Dylan and {Genel}, Shy and {Vogelsberger}, Mark and {Naiman}, Jill and {Torrey}, Paul},
        title = "{Simulating galaxy formation with black hole driven thermal and kinetic feedback}",
      journal = {\mnras},
         year = 2017,
        month = mar,
       volume = {465},
       number = {3},
        pages = {3291-3308},
          doi = {10.1093/mnras/stw2944},
archivePrefix = {arXiv},
       eprint = {1607.03486},
 primaryClass = {astro-ph.GA},
       adsurl = {https://ui.adsabs.harvard.edu/abs/2017MNRAS.465.3291W}
}

@ARTICLE{2009MNRAS.399..497D,
       author = {{Dolag}, K. and {Borgani}, S. and {Murante}, G. and {Springel}, V.},
        title = "{Substructures in hydrodynamical cluster simulations}",
      journal = {\mnras},
         year = 2009,
        month = oct,
       volume = {399},
       number = {2},
        pages = {497-514},
          doi = {10.1111/j.1365-2966.2009.15034.x},
archivePrefix = {arXiv},
       eprint = {0808.3401},
 primaryClass = {astro-ph},
       adsurl = {https://ui.adsabs.harvard.edu/abs/2009MNRAS.399..497D}
}

@ARTICLE{2016A&A...594A..13P,
       author = {{Planck Collaboration} and {Ade}, P.~A.~R. and {Aghanim}, N. et al.},
        title = "{Planck 2015 results. XIII. Cosmological parameters}",
      journal = {\aap},
         year = 2016,
       volume = {594},
          eid = {A13},
        pages = {A13},
         doi = {10.1051/0004-6361/201525830},
archivePrefix = {arXiv},
       eprint = {1502.01589},
 primaryClass = {astro-ph.CO},
       adsurl = {https://ui.adsabs.harvard.edu/abs/2016A&A...594A..13P}
}

@ARTICLE{lu2021hot,
       author = {{Lu}, Shengdong and {Xu}, Dandan and {Wang}, Yunchong and {Chen}, Yanmei and {Zhu}, Ling and {Mao}, Shude and {Springel}, Volker and {Wang}, Jing and {Vogelsberger}, Mark and {Hernquist}, Lars},
        title = "{Hot and counter-rotating star-forming disc galaxies in IllustrisTNG and their real-world counterparts}",
      journal = {\mnras},
         year = 2021,
        month = may,
       volume = {503},
       number = {1},
        pages = {726-742},
          doi = {10.1093/mnras/stab497},
archivePrefix = {arXiv},
       eprint = {2011.01949},
 primaryClass = {astro-ph.GA},
       adsurl = {https://ui.adsabs.harvard.edu/abs/2021MNRAS.503..726L}
}

@ARTICLE{10.1093/mnras/stab3167,
       author = {{Wang}, Sen and {Xu}, Dandan and {Lu}, Shengdong and {Cai}, Zheng and {Xiang}, Maosheng and {Mao}, Shude and {Springel}, Volker and {Hernquist}, Lars},
        title = "{From large-scale environment to CGM angular momentum to star-forming activities - I. Star-forming galaxies}",
      journal = {\mnras},
         year = 2022,
        month = jan,
       volume = {509},
       number = {3},
        pages = {3148-3162},
          doi = {10.1093/mnras/stab3167},
archivePrefix = {arXiv},
       eprint = {2109.06200},
 primaryClass = {astro-ph.GA},
       adsurl = {https://ui.adsabs.harvard.edu/abs/2022MNRAS.509.3148W}
}

@ARTICLE{2022MNRAS.509.2707L,
       author = {{Lu}, Shengdong and {Xu}, Dandan and {Wang}, Sen and {Cai}, Zheng and {He}, Chuan and {Xu}, C. Kevin and {Xia}, Xiaoyang and {Mao}, Shude and {Springel}, Volker and {Hernquist}, Lars},
        title = "{From large-scale environment to CGM angular momentum to star forming activities - II. Quenched galaxies}",
      journal = {\mnras},
         year = 2022,
        month = jan,
       volume = {509},
       number = {2},
        pages = {2707-2719},
          doi = {10.1093/mnras/stab3169},
archivePrefix = {arXiv},
       eprint = {2109.06197},
 primaryClass = {astro-ph.GA},
       adsurl = {https://ui.adsabs.harvard.edu/abs/2022MNRAS.509.2707L}
}

@ARTICLE{benavides2025environmental,
       author = {{Benavides}, Jos{\'e} A. and {Navarro}, Julio F. and {Sales}, Laura V. and {P{\'e}rez}, Isabel and {Bidaran}, Bahar},
        title = "{The Environmental Quenching Mechanisms of Field Dwarf Galaxies}",
      journal = {\apj},
         year = 2025,
        month = may,
       volume = {985},
       number = {1},
          eid = {86},
        pages = {86},
          doi = {10.3847/1538-4357/adced0},
archivePrefix = {arXiv},
       eprint = {2501.13159},
 primaryClass = {astro-ph.GA},
       adsurl = {https://ui.adsabs.harvard.edu/abs/2025ApJ...985...86B}
}

@ARTICLE{10.1093/mnras/stac3282,
       author = {{Herzog}, Georg and {Ben{\'\i}tez-Llambay}, Alejandro and {Fumagalli}, Michele},
        title = "{The present-day gas content of simulated field dwarf galaxies}",
      journal = {\mnras},
         year = 2023,
        month = feb,
       volume = {518},
       number = {4},
        pages = {6305-6317},
          doi = {10.1093/mnras/stac3282},
archivePrefix = {arXiv},
       eprint = {2209.11782},
 primaryClass = {astro-ph.GA},
       adsurl = {https://ui.adsabs.harvard.edu/abs/2023MNRAS.518.6305H}
}

@ARTICLE{2012MNRAS.421.1007K,
       author = {{Kelvin}, Lee S. and {Driver}, Simon P. and {Robotham}, Aaron S.~G. and {Hill}, David T. and {Alpaslan}, Mehmet and {Baldry}, Ivan K. and {Bamford}, Steven P. and {Bland-Hawthorn}, Joss and {Brough}, Sarah and {Graham}, Alister W. and {H{\"a}ussler}, Boris and {Hopkins}, Andrew M. and {Liske}, Jochen and {Loveday}, Jon and {Norberg}, Peder and {Phillipps}, Steven and {Popescu}, Cristina C. and {Prescott}, Matthew and {Taylor}, Edward N. and {Tuffs}, Richard J.},
        title = "{Galaxy And Mass Assembly (GAMA): Structural Investigation of Galaxies via Model Analysis}",
      journal = {\mnras},
         year = 2012,
        month = apr,
       volume = {421},
       number = {2},
        pages = {1007-1039},
          doi = {10.1111/j.1365-2966.2012.20355.x},
archivePrefix = {arXiv},
       eprint = {1112.1956},
 primaryClass = {astro-ph.CO},
       adsurl = {https://ui.adsabs.harvard.edu/abs/2012MNRAS.421.1007K}
}

@ARTICLE{Moustakas_2013,
       author = {{Moustakas}, John and {Coil}, Alison L. and {Aird}, James and {Blanton}, Michael R. and {Cool}, Richard J. and {Eisenstein}, Daniel J. and {Mendez}, Alexander J. and {Wong}, Kenneth C. and {Zhu}, Guangtun and {Arnouts}, St{\'e}phane},
        title = "{PRIMUS: Constraints on Star Formation Quenching and Galaxy Merging, and the Evolution of the Stellar Mass Function from z = 0-1}",
      journal = {\apj},
         year = 2013,
        month = apr,
       volume = {767},
       number = {1},
          eid = {50},
        pages = {50},
          doi = {10.1088/0004-637X/767/1/50},
archivePrefix = {arXiv},
       eprint = {1301.1688},
 primaryClass = {astro-ph.CO},
       adsurl = {https://ui.adsabs.harvard.edu/abs/2013ApJ...767...50M}
}

@ARTICLE{10.1093/mnras/stw2982,
       author = {{Ben{\'\i}tez-Llambay}, Alejandro and {Navarro}, Julio F. and {Frenk}, Carlos S. and {Sawala}, Till and {Oman}, Kyle and {Fattahi}, Azadeh and {Schaller}, Matthieu and {Schaye}, Joop and {Crain}, Robert A. and {Theuns}, Tom},
        title = "{The properties of `dark' {\ensuremath{\Lambda}}CDM haloes in the Local Group}",
      journal = {\mnras},
         year = 2017,
        month = mar,
       volume = {465},
       number = {4},
        pages = {3913-3926},
          doi = {10.1093/mnras/stw2982},
archivePrefix = {arXiv},
       eprint = {1609.01301},
 primaryClass = {astro-ph.GA},
       adsurl = {https://ui.adsabs.harvard.edu/abs/2017MNRAS.465.3913B}
}

@ARTICLE{Weisz_2014,
       author = {{Weisz}, Daniel R. and {Dolphin}, Andrew E. and {Skillman}, Evan D. and {Holtzman}, Jon and {Gilbert}, Karoline M. and {Dalcanton}, Julianne J. and {Williams}, Benjamin F.},
        title = "{The Star Formation Histories of Local Group Dwarf Galaxies. I. Hubble Space Telescope/Wide Field Planetary Camera 2 Observations}",
      journal = {\apj},
         year = 2014,
        month = jul,
       volume = {789},
       number = {2},
          eid = {147},
        pages = {147},
          doi = {10.1088/0004-637X/789/2/147},
archivePrefix = {arXiv},
       eprint = {1404.7144},
 primaryClass = {astro-ph.GA},
       adsurl = {https://ui.adsabs.harvard.edu/abs/2014ApJ...789..147W}
}

@ARTICLE{Gallart_2015,
       author = {{Gallart}, Carme and {Monelli}, Matteo and {Mayer}, Lucio and {Aparicio}, Antonio and {Battaglia}, Giuseppina and {Bernard}, Edouard J. and {Cassisi}, Santi and {Cole}, Andrew A. and {Dolphin}, Andrew E. and {Drozdovsky}, Igor and {Hidalgo}, Sebastian L. and {Navarro}, Julio F. and {Salvadori}, Stefania and {Skillman}, Evan D. and {Stetson}, Peter B. and {Weisz}, Daniel R.},
        title = "{The ACS LCID Project: On the Origin of Dwarf Galaxy Types{\textemdash}A Manifestation of the Halo Assembly Bias?}",
      journal = {\apjl},
         year = 2015,
        month = oct,
       volume = {811},
       number = {2},
          eid = {L18},
        pages = {L18},
          doi = {10.1088/2041-8205/811/2/L18},
archivePrefix = {arXiv},
       eprint = {1507.08350},
 primaryClass = {astro-ph.GA},
       adsurl = {https://ui.adsabs.harvard.edu/abs/2015ApJ...811L..18G}
}

@ARTICLE{1972ApJ...176....1G,
       author = {{Gunn}, James E. and {Gott}, III, J. Richard},
        title = "{On the Infall of Matter Into Clusters of Galaxies and Some Effects on Their Evolution}",
      journal = {\apj},
         year = 1972,
        month = aug,
       volume = {176},
        pages = {1},
          doi = {10.1086/151605},
       adsurl = {https://ui.adsabs.harvard.edu/abs/1972ApJ...176....1G}
}

@ARTICLE{10.1046/j.1365-8711.1999.02715.x,
       author = {{Abadi}, Mario G. and {Moore}, Ben and {Bower}, Richard G.},
        title = "{Ram pressure stripping of spiral galaxies in clusters}",
      journal = {\mnras},
         year = 1999,
        month = oct,
       volume = {308},
       number = {4},
        pages = {947-954},
          doi = {10.1046/j.1365-8711.1999.02715.x},
archivePrefix = {arXiv},
       eprint = {astro-ph/9903436},
 primaryClass = {astro-ph},
       adsurl = {https://ui.adsabs.harvard.edu/abs/1999MNRAS.308..947A}
}

@ARTICLE{10.1111/j.1365-2966.2012.21188.x,
       author = {{Wetzel}, Andrew R. and {Tinker}, Jeremy L. and {Conroy}, Charlie},
        title = "{Galaxy evolution in groups and clusters: star formation rates, red sequence fractions and the persistent bimodality}",
      journal = {\mnras},
         year = 2012,
        month = jul,
       volume = {424},
       number = {1},
        pages = {232-243},
          doi = {10.1111/j.1365-2966.2012.21188.x},
archivePrefix = {arXiv},
       eprint = {1107.5311},
 primaryClass = {astro-ph.CO},
       adsurl = {https://ui.adsabs.harvard.edu/abs/2012MNRAS.424..232W}
}

@ARTICLE{10.1093/mnras/stad2576,
       author = {{Samuel}, Jenna and {Pardasani}, Bhavya and {Wetzel}, Andrew and {Santistevan}, Isaiah and {Boylan-Kolchin}, Michael and {Moreno}, Jorge and {Faucher-Gigu{\`e}re}, Claude-Andr{\'e}},
        title = "{A jolt to the system: ram pressure on low-mass galaxies in simulations of the Local Group}",
      journal = {\mnras},
         year = 2023,
        month = nov,
       volume = {525},
       number = {3},
        pages = {3849-3864},
          doi = {10.1093/mnras/stad2576},
archivePrefix = {arXiv},
       eprint = {2212.07518},
 primaryClass = {astro-ph.GA},
       adsurl = {https://ui.adsabs.harvard.edu/abs/2023MNRAS.525.3849S}
}

@ARTICLE{Geha_2012,
       author = {{Geha}, M. and {Blanton}, M.~R. and {Yan}, R. and {Tinker}, J.~L.},
        title = "{A Stellar Mass Threshold for Quenching of Field Galaxies}",
      journal = {\apj},
         year = 2012,
        month = sep,
       volume = {757},
       number = {1},
          eid = {85},
        pages = {85},
          doi = {10.1088/0004-637X/757/1/85},
archivePrefix = {arXiv},
       eprint = {1206.3573},
 primaryClass = {astro-ph.CO},
       adsurl = {https://ui.adsabs.harvard.edu/abs/2012ApJ...757...85G}
}

@ARTICLE{10.1093/mnras/stz2338,
       author = {{Pillepich}, Annalisa and {Nelson}, Dylan and {Springel}, Volker and {Pakmor}, R{\"u}diger and {Torrey}, Paul and {Weinberger}, Rainer and {Vogelsberger}, Mark and {Marinacci}, Federico and {Genel}, Shy and {van der Wel}, Arjen and {Hernquist}, Lars},
        title = "{First results from the TNG50 simulation: the evolution of stellar and gaseous discs across cosmic time}",
      journal = {\mnras},
         year = 2019,
        month = dec,
       volume = {490},
       number = {3},
        pages = {3196-3233},
          doi = {10.1093/mnras/stz2338},
archivePrefix = {arXiv},
       eprint = {1902.05553},
 primaryClass = {astro-ph.GA},
       adsurl = {https://ui.adsabs.harvard.edu/abs/2019MNRAS.490.3196P}
}

@ARTICLE{Pillepich2018,
       author = {{Pillepich}, Annalisa and {Springel}, Volker and {Nelson}, Dylan and {Genel}, Shy and {Naiman}, Jill and {Pakmor}, R{\"u}diger and {Hernquist}, Lars and {Torrey}, Paul and {Vogelsberger}, Mark and {Weinberger}, Rainer and {Marinacci}, Federico},
        title = "{Simulating galaxy formation with the IllustrisTNG model}",
      journal = {\mnras},
         year = 2018,
        month = jan,
       volume = {473},
       number = {3},
        pages = {4077-4106},
          doi = {10.1093/mnras/stx2656},
archivePrefix = {arXiv},
       eprint = {1703.02970},
 primaryClass = {astro-ph.GA},
       adsurl = {https://ui.adsabs.harvard.edu/abs/2018MNRAS.473.4077P}
}

@ARTICLE{Wright2024,
       author = {{Wright}, Ruby J. and {Somerville}, Rachel S. and {Lagos}, Claudia del P. and {Schaller}, Matthieu and {Dav{\'e}}, Romeel and {Angl{\'e}s-Alc{\'a}zar}, Daniel and {Genel}, Shy},
        title = "{The baryon cycle in modern cosmological hydrodynamical simulations}",
      journal = {\mnras},
         year = 2024,
        month = aug,
       volume = {532},
       number = {3},
        pages = {3417-3440},
          doi = {10.1093/mnras/stae1688},
archivePrefix = {arXiv},
       eprint = {2402.08408},
 primaryClass = {astro-ph.GA},
       adsurl = {https://ui.adsabs.harvard.edu/abs/2024MNRAS.532.3417W}
}

@ARTICLE{weinberger2018supermassive,
       author = {{Weinberger}, Rainer and {Springel}, Volker and {Pakmor}, R{\"u}diger and {Nelson}, Dylan and {Genel}, Shy and {Pillepich}, Annalisa and {Vogelsberger}, Mark and {Marinacci}, Federico and {Naiman}, Jill and {Torrey}, Paul and {Hernquist}, Lars},
        title = "{Supermassive black holes and their feedback effects in the IllustrisTNG simulation}",
      journal = {\mnras},
         year = 2018,
        month = sep,
       volume = {479},
       number = {3},
        pages = {4056-4072},
          doi = {10.1093/mnras/sty1733},
archivePrefix = {arXiv},
       eprint = {1710.04659},
 primaryClass = {astro-ph.GA},
       adsurl = {https://ui.adsabs.harvard.edu/abs/2018MNRAS.479.4056W}
}

@ARTICLE{mamon2004origin,
       author = {{Mamon}, G.~A. and {Sanchis}, T. and {Salvador-Sol{\'e}}, E. and {Solanes}, J.~M.},
        title = "{The origin of H I-deficiency in galaxies on the outskirts of the Virgo cluster. I. How far can galaxies bounce out of clusters?}",
      journal = {\aap},
         year = 2004,
        month = feb,
       volume = {414},
        pages = {445-451},
          doi = {10.1051/0004-6361:20034155},
archivePrefix = {arXiv},
       eprint = {astro-ph/0310709},
 primaryClass = {astro-ph},
       adsurl = {https://ui.adsabs.harvard.edu/abs/2004A&A...414..445M}
}

@ARTICLE{Bidaran_2025,
       author = {{Bidaran}, Bahar and {P{\'e}rez}, Isabel and {S{\'a}nchez-Menguiano}, Laura and {Argudo-Fern{\'a}ndez}, Mar{\'\i}a and {Ferr{\'e}-Mateu}, Anna and {Navarro}, Julio F. and {Peletier}, Reynier F. and {Ruiz-Lara}, Tom{\'a}s and {van de Ven}, Glenn and {Verley}, Simon and {Zurita}, Almudena and {Duarte Puertas}, Salvador and {Falc{\'o}n-Barroso}, Jes{\'u}s and {S{\'a}nchez-Bl{\'a}zquez}, Patricia and {Jim{\'e}nez}, Andoni},
        title = "{The puzzle of isolated and quenched dwarf galaxies in cosmic voids}",
      journal = {\aap},
         year = 2025,
        month = jan,
       volume = {693},
          eid = {L16},
        pages = {L16},
          doi = {10.1051/0004-6361/202452688},
archivePrefix = {arXiv},
       eprint = {2501.02910},
 primaryClass = {astro-ph.GA},
       adsurl = {https://ui.adsabs.harvard.edu/abs/2025A&A...693L..16B}
}

@ARTICLE{rey2020edge,
       author = {{Rey}, Martin P. and {Pontzen}, Andrew and {Agertz}, Oscar and {Orkney}, Matthew D.~A. and {Read}, Justin I. and {Rosdahl}, Joakim},
        title = "{EDGE: from quiescent to gas-rich to star-forming low-mass dwarf galaxies}",
      journal = {\mnras},
         year = 2020,
        month = sep,
       volume = {497},
       number = {2},
        pages = {1508-1520},
          doi = {10.1093/mnras/staa1640},
archivePrefix = {arXiv},
       eprint = {2004.09530},
 primaryClass = {astro-ph.GA},
       adsurl = {https://ui.adsabs.harvard.edu/abs/2020MNRAS.497.1508R}
}

@ARTICLE{10.1093/mnras/256.1.43P,
       author = {{Efstathiou}, G.},
        title = "{Suppressing the formation of dwarf galaxies via photoionization}",
      journal = {\mnras},
         year = 1992,
        month = may,
       volume = {256},
       number = {2},
        pages = {43P-47P},
          doi = {10.1093/mnras/256.1.43P},
       adsurl = {https://ui.adsabs.harvard.edu/abs/1992MNRAS.256P..43E}
}

@ARTICLE{hopkins2014galaxies,
       author = {{Hopkins}, Philip F. and {Kere{\v{s}}}, Du{\v{s}}an and {O{\~n}orbe}, Jos{\'e} and {Faucher-Gigu{\`e}re}, Claude-Andr{\'e} and {Quataert}, Eliot and {Murray}, Norman and {Bullock}, James S.},
        title = "{Galaxies on FIRE (Feedback In Realistic Environments): stellar feedback explains cosmologically inefficient star formation}",
      journal = {\mnras},
         year = 2014,
        month = nov,
       volume = {445},
       number = {1},
        pages = {581-603},
          doi = {10.1093/mnras/stu1738},
archivePrefix = {arXiv},
       eprint = {1311.2073},
 primaryClass = {astro-ph.CO},
       adsurl = {https://ui.adsabs.harvard.edu/abs/2014MNRAS.445..581H}
}

@ARTICLE{stilp2013global,
       author = {{Stilp}, Adrienne M. and {Dalcanton}, Julianne J. and {Warren}, Steven R. and {Skillman}, Evan and {Ott}, J{\"u}rgen and {Koribalski}, B{\"a}rbel},
        title = "{Global H I Kinematics in Dwarf Galaxies}",
      journal = {\apj},
         year = 2013,
        month = mar,
       volume = {765},
       number = {2},
          eid = {136},
        pages = {136},
          doi = {10.1088/0004-637X/765/2/136},
archivePrefix = {arXiv},
       eprint = {1301.1989},
 primaryClass = {astro-ph.GA},
       adsurl = {https://ui.adsabs.harvard.edu/abs/2013ApJ...765..136S}
}

@ARTICLE{tamburro2009driving,
       author = {{Tamburro}, D. and {Rix}, H.-W. and {Leroy}, A.~K. and {Mac Low}, M.-M. and {Walter}, F. and {Kennicutt}, R.~C. and {Brinks}, E. and {de Blok}, W.~J.~G.},
        title = "{What is Driving the H I Velocity Dispersion?}",
      journal = {\aj},
         year = 2009,
        month = may,
       volume = {137},
       number = {5},
        pages = {4424-4435},
          doi = {10.1088/0004-6256/137/5/4424},
archivePrefix = {arXiv},
       eprint = {0903.0183},
 primaryClass = {astro-ph.GA},
       adsurl = {https://ui.adsabs.harvard.edu/abs/2009AJ....137.4424T}
}

@ARTICLE{wang2024larger,
       author = {{Wang}, Sen and {Xu}, Dandan and {Lu}, Shengdong},
        title = "{From Larger-scale Cold-gas Angular Momentum Environments to Galaxy Star Formation Activity}",
      journal = {\apj},
         year = 2025,
        month = jun,
       volume = {986},
       number = {1},
          eid = {85},
        pages = {85},
          doi = {10.3847/1538-4357/add155},
archivePrefix = {arXiv},
       eprint = {2411.16849},
 primaryClass = {astro-ph.GA},
       adsurl = {https://ui.adsabs.harvard.edu/abs/2025ApJ...986...85W}
}

@ARTICLE{2025OJAp....8E..43B,
       author = {{Bhattacharyya}, Joy and {Peter}, Annika H.~G. and {Leauthaud}, Alexie},
        title = "{Dwarf Galaxies in the TNG50 Field: connecting their Star-formation Rates with their Environments}",
      journal = {The Open Journal of Astrophysics},
         year = 2025,
        month = apr,
       volume = {8},
          eid = {43},
        pages = {43},
          doi = {10.33232/001c.137130},
archivePrefix = {arXiv},
       eprint = {2501.01946},
 primaryClass = {astro-ph.GA},
       adsurl = {https://ui.adsabs.harvard.edu/abs/2025OJAp....8E..43B}
}

@ARTICLE{2002AJ....124.2639V,
       author = {{van der Marel}, Roeland P. and {Alves}, David R. and {Hardy}, Eduardo and {Suntzeff}, Nicholas B.},
        title = "{New Understanding of Large Magellanic Cloud Structure, Dynamics, and Orbit from Carbon Star Kinematics}",
      journal = {\aj},
         year = 2002,
        month = nov,
       volume = {124},
       number = {5},
        pages = {2639-2663},
          doi = {10.1086/343775},
archivePrefix = {arXiv},
       eprint = {astro-ph/0205161},
 primaryClass = {astro-ph},
       adsurl = {https://ui.adsabs.harvard.edu/abs/2002AJ....124.2639V}
}

\end{document}